\documentclass[12pt,letterpaper]{article}
\usepackage{jcappub}
\usepackage{amsmath,amssymb,scalefnt}
\newcommand{\beq}{\begin{equation}}
\newcommand{\eeq}{\end{equation}}
\newcommand{\bea}{\begin{eqnarray}}
\newcommand{\eea}{\end{eqnarray}}
\def\pa{\partial}
\newcommand{\vev}[1]{\left\langle#1\right\rangle}
\newcommand{\eqn}[1]{Eq.~(\ref{#1})}
\newcommand{\eqns}[2]{Eqs.~(\ref{#1}),(\ref{#2})}

\def\GeV{\hbox{GeV}}

\newcommand{\reference}[1]{Ref.~\cite{#1}}
\newcommand{\abbrev}{\scalefont{.9}}
\newcommand{\kahler}{K\"{a}hler\ }

\newcommand{\sic}{supersymmetric}

\newcommand{\sy}{{supersymmetry}}
\newcommand{\sm}{{\abbrev SM}}
\newcommand{\gut}{{\abbrev GUT}}
\newcommand{\guts}{{\abbrev GUTS}}
\newcommand{\mssm}{{\abbrev MSSM}}
\newcommand{\nmssm}{{\abbrev NMSSM}}

\newcommand{\half}{\frac{1}{2}}
\newcommand{\la}{\lambda}
\newcommand{\labar}{\bar\lambda}
\newcommand{\Vbar}{\bar V}
\newcommand{\xbar}{\bar x}
\newcommand{\Ttil}{\tilde T}
\newcommand{\nn}{\nonumber\\}
\def\Tr{{\rm Tr }}

\def\Abar{\overline{A}}
\def\Ybar{\overline{Y}}
\title{GUT Scalar Potentials for Higgs Inflation}
\author[a,1]{Martin B Einhorn,%
\note{Emeritus Professor, University of Michigan}}
\author[b]{D R Timothy Jones}
\affiliation[a]{Kavli Institute for Theoretical Physics,\\
University of California, Santa Barbara CA 93106-4030, USA}%
\affiliation[b]{Dept. of Mathematical Sciences,\\
University of Liverpool, Liverpool L69 3BX, UK}
\emailAdd{meinhorn@umich.edu}
\emailAdd{drtj@liv.ac.uk}

\abstract{

Motivated by the idea that there is new physics beyond the Standard
Model (\sm), we have investigated 
a number of models for Grand Unified Theories (\guts) in
four dimensions for the possibility that their
Higgs fields might be responsible for inflation in the
early universe.  
In addition to models having an intrinsic
Planck mass parameter, we have entertained classically scale invariant
models in which the Planck scale itself as well as the \gut\ scale is
induced by  spontaneous breaking of the gauge symmetry.  We found that
in non-\sic\ $SU(5)$ with the usual Higgs in the adjoint
representation but with large non-minimal coupling to the curvature,
there appear to be several possible flat directions that might lead to
inflation.  Interestingly, the one of lowest energy is the breaking into
$SU(3)\otimes SU(2)\otimes U(1)$ that is suggested by gauge coupling
unification.  Further, we show that this flat direction is stable
against small fluctuations in other directions.

{\hskip 0.25in} We attempted to extend this to similar \sic\ \guts, both
global and supergravity, but did not succeed in finding a
phenomenologically acceptable model of this type.  As is often the case,
such models suffered either from a negative vacuum energy or from
tachyonic modes. We also considered a variant of an ``inverted
hierarchy" model in which  the \gut\ scale is set by dimensional
transmutation, but were unable to find a phenomenologically
acceptable model.
}

\arxivnumber{1207.1710 (preprint:NSF-KITP-12-102, LTH 947)}

\begin{document}
\maketitle

\section{Introduction}

Quantum field theories in curved backgrounds (or with quantised gravity) in 
four dimensions incorporating scalar fields may contain terms of the generic form
\beq
{\cal L} \supset \sqrt{g} \xi \phi^2 R 
\label{eq:bsterma}
\eeq
where $\xi$ is a dimensionless coefficient. 

In non-\sic\  theories, $\xi$ is renormalised; with $\xi = 0$ {\it not\/}
being a fixed point with respect  to the renormalisation group.   There
has been recent interest~\cite{Bezrukov:2007ep}-\cite{Bezrukov:2010jz}\  in
the possibility that the  term of this form involving the Higgs boson 
which is permitted in the Standard Model might suffice  to allow the
Higgs to be the inflaton. We shall refer to this general 
paradigm as the Bezrukov-Shaposhnikov (BS) scenario.  
The attraction of this concept is obvious;
instead of the ad hoc introduction of a  new scalar sector specifically
engineered to produce an inflationary era,  we have a sector already
conceived of for other (excellent) reasons  performing the same
function. The point is that for field values $\vev\phi$ such that 
\beq \xi \vev\phi^2 \gtrsim M_P^2, \label{eq:xibiga} 
\eeq
the scalar potential (in the Jordan frame) is nearly flat and amenable to 
slow-roll inflation. Crucially, it is possible to entertain the possibility 
$\xi >> 1$ without (obviously) violating perturbation theory. 
There has been vigorous debate over the range in field values 
(of the inflaton) for which the classical field analysis remains valid 
as an effective field theory; a particularly detailed analysis appears in 
\reference{Bezrukov:2010jz} (see also \reference{Lerner:2011it}). An interesting question is whether 
rendering the theory supersymmetric has any impact here; but we 
will not address this issue in this paper. 

Here we explore the possibility that the inflaton might again be a Higgs, but 
associated  with the spontaneous breaking of a \gut\ rather than the
electroweak gauge group; we have in mind in particular the case of a 24
of $SU(5)$; (a case considered some time ago, in fact, by Salopek et al 
in \reference{Salopek:1988qh}).     We also discuss the (even more) radical  possibility that
\eqn{eq:bsterma}\ is in fact the {\it only\/} term linear in the
curvature  in the action; that is to say, the usual Einstein-Hilbert term $M_P^2 R$  
(or its \sic\ generalisation) is absent, and the classical
theory is scale invariant. We will explore the ramifications of this general approach 
elsewhere\footnote{For other previous work in this general direction, see 
for example~\reference{Zee:1978wi}-\reference{grs}.};
noting, however, that, as emphasised in \reference{Salopek:1988qh}), there 
are then difficulties with the interpretation of the field associated with the Higgs flat direction as the 
inflaton since it decouples from the matter fields and hence does not contribute to reheating.

We also generalise to the \sic\ case, 
which was first considered in 
\reference{Einhorn:2009bh}, and further developed in 
Refs.~\cite{Ferrara:2010yw}-\cite{Pallis:2011gr}. 
It turns out to be not possible to implement the idea (a Higgs inflaton) in its basic form 
in either the \mssm~\cite{Einhorn:2009bh}, or the \nmssm~\cite{Ferrara:2010yw}, because 
the candidate flat directions suffer from tachyonic instabilities. It is possible to circumvent 
this problem by generalising to a non-minimal K\" ahler potential; we will return to this issue, 
but in this paper we prefer to concentrate on the minimal case when 
the K\"ahler potential is augmented only by terms characterised by 
{\it dimensionless\/} coupling constants.  
We discuss the form of the scalar potential first in a general class of \sic\ models, 
eventually specialising once again to the case of an adjoint field. 
We present a particularly simple form for the scalar potential
both in general  and in the special case of $SU(N)$. 
We also consider in detail the more complicated example of Witten's "Inverted Hierarchy" model, 
which involves two chiral adjoint multiplets. In the supersymmetric 
cases that we explored, we have typically encountered 
the same instability problem (alluded to above) 
associated with the \mssm\ and the \nmssm, but we have 
not performed a completely general analysis 
allowing for extrema with complex field values.

\section{The adjoint case for non-\sic\ $SU(5)$}

Let us begin more generally with $SU(N)$. For a single hermitian adjoint
multiplet, the most general possible quartic scale invariant potential in the Jordan frame is
\beq
V_J (\Phi) = \lambda_1 \Tr \Phi^4 + \lambda_2 (\Tr \Phi^2)^2,
\eeq
where $\Phi = \lambda^A \phi^A/\sqrt{2}$, and the $\phi^A$ are real. 
Some results from group theory and our notational 
conventions are to be found in Appendix~\ref{group}.
 
If we include a term $\xi \Tr \Phi^2 R$ in the Lagrangian in the Jordan frame, then 
in the Einstein frame, we have (in the usual case when the Lagrangian also contains an 
Einstein term $M_P^2 R$) the potential~\cite{Bezrukov:2007ep} 
\beq V (\Phi) = \left(\frac{M_P^2}{X}\right)^2 V_J(\phi)
\label{vadj}
\eeq 
with $X = M_P^2 + \xi \Tr \Phi^2$. In the {\it scale invariant\/} case (when the $M_P^2 R$ term is 
absent) we have instead   
$X = \xi \Tr \Phi^2$, where $M_P$ is now an {\it arbitrary\/} scale 
introduced in the course of performing the 
conformal transformation which connects the two frames. 

As we indicated earlier, his model was in fact considered in 
\reference{Salopek:1988qh}; the general analysis of the potential 
which appears below is, however, new.  

Now a hermitian matrix can be rendered real and diagonal by means of a
unitary transformation. Since $V$ is invariant under an arbitrary 
unitary transformation, 
in order to find the extrema of the potential in
the Einstein frame  it will suffice to consider the case
\beq \Phi = \hbox{diag}\, (x_1\cdots x_i \cdots
x_N) \quad\hbox{and traceless}, \label{phitracec}
\eeq
with $x_i$ real.

Let us first analyse the scale invariant case. We find that  the $x_i$ satisfy
the equation
\beq x^3 - \frac{T_4}{T_2}x - \frac{1}{N}T_3 = 0, 
\label{extnoscaleb}
\eeq
where $T_m = \Tr \Phi^m$.
Of course $V$ is not well defined at $\Phi = 0$.
Since \eqn{extnoscaleb} is a cubic, there are at most three distinct
solutions for $x_i.$ For $N$ even, there is the obvious solution  
\beq
\Phi = \Lambda\, \hbox{diag}\,  (1,-1,\cdots 1,-1), 
\label{eq:Nevenb}
\eeq 
with  \beq V = \frac{M_P^4}{4\xi^2}\left( \lambda_2 +
\frac{\lambda_1}{N}\right) \eeq This flat direction represents, in fact,
the minimum of $V$ for all even $N$.

For odd $N$ the situation is more complicated; let us turn to  the $SU(5)$ case.
Suppose at least one of the $x_i$ is zero.
It follows from \eqn{extnoscaleb}\ that
$T_3 = 0$. It is therefore easy to see that the possible solutions
for $SU(5)$ are $\Phi = \hbox{diag}\,  (1,-1,0,0,0)$,
$\Phi = \hbox{diag}\,  (1,-1,1,-1,0)$.
with results for $V$ of
\beq 
V = \frac{M_P^4}{4\xi^2}\left( \lambda_2 + \frac{\lambda_1}{2}\right)
\eeq
and
\beq
V = \frac{M_P^4}{4\xi^2}\left( \lambda_2 + \frac{\lambda_1}{4}\right)   
\eeq
respectively.

If none of the $x_i$ are zero then we have the possible forms
\bea
A: \Phi &=& \hbox{diag}\,  (1,1,1,z,-3-z)\\
B: \Phi &=& \hbox{diag}\,  (1,1,z,z,-2-2z).
\eea
Substitution of these forms in \eqn{vadj}\ and plotting $V$
as a function of $z$ (or seeking consistent solutions
for all $x_i$ of \eqn{extnoscaleb}) reveals that there are in fact
solutions corresponding to
\bea
A: \Phi &=& \hbox{diag}\,  (1,1,1,-3/2,-3/2)\\
B: \Phi &=& \hbox{diag}\,  (1,1,1,1,-4)
\eea
with
\beq
V = \frac{M_P^4}{4\xi^2}\left( \lambda_2 + \frac{7\lambda_1}{30}\right)
\label{eq:vresult321}
\eeq
and
\beq
V = \frac{M_P^4}{4\xi^2}\left( \lambda_2 + \frac{13\lambda_1}{20}\right)
\eeq
respectively.

Note that the $SU(3)\otimes SU(2) \otimes U(1)$ solution  has the
smallest energy. It is in fact stable against all quadratic
fluctuations, as shown in Appendix~\ref{sec:stability}!

Let us turn now to analyse the potential {\it with\/} the Einstein term, 
$X = M_P^2 + \xi T_2$. 
Then we find (setting $M_P = 1$) that for an extremum the $x_i$ must satisfy 
\beq 
\la_1(1+\xi T_2) x^3 + (\la_2 T_2 - \la_1 \xi T_4)x - \frac{1}{N}\xi \la_1 T_2 T_3 = 0.
\label{extscale}
\eeq
It is easy to see that \eqn{extscale}\ reduces to \eqn{extnoscaleb}\ in the limit $\xi \to \infty$. 
However, from \eqn{extscale}\ it follows at once that 
\beq 
V_J (\Phi) = \la_1 T_4 + \la_2 T_2^2 = 0.
\eeq
Thus in the presence of the Einstein term, the only true extrema of the potential 
\beq 
V = \frac{V_J (\Phi)}{(1 + \xi T_2)^2}
\label{vjpot}
\eeq
have $V = 0$. 

This result is perfectly consistent with an inflationary interpretation; at large 
$\xi T_2$, the flat direction of lowest energy is 
the one corresponding to a $SU(3) \otimes SU(2) \otimes U(1)$ vacuum, with 
slow-roll towards the true minimum at $\Phi = 0$. 

Note: we can understand the extremum of \eqn{vjpot} in a simple way as follows. 
Suppose $V$ as defined in \eqn{vjpot} has an extremum $V = \Vbar$ for $x_i = \xbar_i$. 
Then consider $\Vbar_{\lambda} = V(\lambda \xbar_i)$.
Evidently
\beq
\Vbar_{\lambda} = (1+\xi T_2)^2 \Vbar \left[\frac{\lambda^4}{(1+\la^2\xi T_2)^2}\right],
\eeq
and from the fact that the function $y = x^4/(1+ax^2)^2$ has, for $a > 0$,  a unique extremum (for finite $x$) at 
$x = 0$ the result follows. 

We now generalise $V_J (\phi)$ by including a mass term as follows:
\beq 
V_J (\Phi) = -m^2 T_2 + \la_1 T_4 + \la_2 T_2^2 = 0.
\eeq
We assume $m^2 << M_P^2$. For field values $\xi T_2 \ll M_P^2$,  
we can ignore the $\xi$-term and the minimisation of 
this potential is an old problem; according to Li~\cite{Li:1973mq}, for 
\beq
\la_2 > 0\quad\hbox{and}\quad \la_2 > -\frac{7}{30}\la_1
\label{eq:lires}
\eeq
the minimum corresponds to 
breaking to $SU(3)\otimes SU(2) \otimes U(1)$, with 
\beq
\vev \Phi = v_{\Phi}\hbox{diag} (2,2,2,-3,-3)
\eeq and 
\beq
v_{\Phi}^2 = \frac{m^2}{60\la_2 + 14 \la_1}.
\eeq
Notice that given \eqn{eq:lires}, 
at least for large scales, we have both $v_{\Phi}^2 > 0$, and $V >0$ 
in \eqn{eq:vresult321}. 

Of course for $|\vev \Phi| \gg v_{\Phi}$ one should account for the 
running of $m^2$ and $\la_{1,2}$ between the two scales; but,  given
\eqn{eq:lires}, it seems natural that  the inflationary era described by
\eqn{eq:vresult321}\ would terminate with a transition to the broken
vacuum. So if we substitute in the original  Jordan frame Lagrangian
\beq 
{\cal L} = \frac{1}{2}(M_P^2 
+ \xi \Tr \Phi^2)R + \frac{1}{2}\Tr ( \pa_{\mu}\Phi \pa^{\mu}\Phi) - V_J (\Phi)
\eeq
the form 
\beq
\Phi = \frac{h}{\sqrt{30}} \hbox{diag} (2,2,2,-3,-3)
\eeq
we obtain, in the first approximation, precisely the results of \cite{Bezrukov:2007ep}\ for the 
slow-roll parameters, namely (setting $M_P = 1$)
\bea
\epsilon &=& \frac{4}{3\xi^2 h^4},\\
N &=& \frac{3\xi (h^2 -h_{\rm end}^2)}{4},\\
\eta &=& -\frac{4}{3\xi h^2}. 
\eea
with $\epsilon \approx 1$ for $h_{end} \approx 1.07 M_P / \sqrt{\xi}$. 
Thus for $\xi \sim 10^6$, inflation can terminate
naturally with a transition to the broken vacuum at a scale $h_{\rm end} \sim 10^{16} \GeV$, while 
$h_0 \sim 9 M_P / \sqrt{\xi} \sim 10^{17}\GeV$. 

Thus the simplest $SU(5)$ model (the original Georgi-Glashow
model~\cite{Georgi:1974sy}) is compatible with the BS inflation
scenario: a Higgs inflaton. This model is, however,  generally regarded
as unsatisfactory for reasons both  theoretical and experimental; most
particularly the increasingly precise limit on  the proton lifetime. 
Note, however, that we may anticipate that our scenario exists in a 
considerable class of $SU(5)$ models, designed, for example, to
alleviate  the doublet-triplet splitting problem, or to increase
somewhat the  unification mass so as to reduce the proton decay rate.
For some recent examples see~\cite{Feldmann:2010yp}-\cite{Schnitter:2012bz}.
Thus it does seem to us that  this formulation of Higgs inflation is of
interest. 

We turn now to the \sic\ case; more popular than  the
non-\sic\ case for reasons which are well known; though not without  its
own problems, regarding, for example, dimension 5 contributions to 
proton decay which are of course absent in the non-\sic\ case.  We will
find that we are unable to construct a simple model with a stable 
trajectory of the kind we have identified above.

\section{The \sic\ case}

Let us  consider the form of the scalar potential in general gauge
invariant  $N=1$ supergravity models with chiral supermultiplets; and
with a  K\"ahler potential modified (in the Einstein frame) by the 
inclusion of a term which in the Jordan frame corresponds to adding to
the  Lagrangian the term  \beq {\cal L} \supset -6 \int\, d^2\Theta\,
{\cal E} X(\Phi) R + \hbox{c.c.,} \label{eq:bsterm} \eeq where $X(\Phi)$
is {\it quadratic\/} in the chiral supermultiplet $\Phi$.  Such a term
is a natural \sic\ generalisation of \eqn{eq:bsterma}. One can of course
entertain the possibility that $X$ contains higher  powers of $\Phi$;
and indeed these may be used to  mitigate the tachyonic instabilities 
which we will encounter. For a viable model of this type, see for
example  \reference{Lee:2010hj}. However  doing so does call into
question the generality of any conclusions reached  against
contributions from yet other higher dimension operators added to the
\kahler  potential or superpotential.

We will find that, generally speaking, for theories with or without the 
$M_P^2 R$ there exist  natural flat directions that are candidates for
slow-roll inflationary eras; but that typically these  suffer from
unstable directions in the complete field space.\footnote{For the
original \nmssm-based model of \reference{Einhorn:2009bh} this was
pointed out  in \reference{Ferrara:2010yw}.} In spite of this rather
negative conclusion,  we believe our discussion remains of interest; for example we 
present a particularly simple form for the scalar potential in a wide class of theories.

\section{The scalar potential}

The scalar part of the $N=1$ supergravity Lagrangian is 
given (in the Einstein frame) by 
\beq 
{\cal L} = g_a{}^b g^{\mu\nu}D_{\mu}\phi^a D_{\nu}\phi_b^* -V(\phi,\phi^*)
\label{eq:scalarL}
\eeq
where $\phi^a$ is the chiral scalar multiplet in an arbitrary representation, 
and $g_a{}^b$ is the \kahler metric, 
given by 
\beq
g_a{}^b = \frac{\pa^2 K}{\pa\phi^a\pa\phi^*_b},
\eeq
where $K(\phi,\phi^*)$ is the \kahler potential. 

The scalar potential $V$ is 
\beq
V = V_F + V_D,
\eeq
where 
\beq
V_D = \frac{1}{2}g^2 \hbox{Re}\, f_{AB} K_a (R^A)^a{}_b \phi^b K_c (R^B)^c{}_d \phi^d
\eeq
and 
\beq
V_F= e^K\Big[(g^{-1})_a{}^b (D^a W^*) D_b W -3W W^*\Big]
\label{eq:effpot}
\eeq
where 
\beq
D_b W\equiv W_b+K_b W,~{\rm and}~W_b\equiv\frac{\partial W}{\partial\phi^b},
\eeq
and $W(\phi^a)$ is the superpotential. Note that 
\beq
D^a W^* = \frac{\pa W^*}{\partial \phi^*_a} 
+ \frac{\pa K}{\partial \phi^*_a}W^* = (D_a W)^*.  
\eeq
Consider the \kahler potential defined by 
\beq
K=-3M_P^2\log[|\Omega(\phi^a,\phi_b^*)|/3],
\label{eq:kform}
\eeq 
with 
\beq
\Omega= \frac{1}{M_P^2}\left[{\textstyle\sum}\phi_a^*\phi^a
-\half\xi(c_{ab}\phi^a\phi^b+c.c.) 
-3 M_P^2\right].
\eeq
where $c_{ab}$  is an invariant tensor of the gauge group. Of course
$c_{ab}$ exists only for certain representations.  For example, in
$SU(5)$ with two Higgs multiplets, one $(H_{u})$ in the ${\bf 5}$ and 
the other $(H_{d})$ in the $({\bf\overline{5}})$ or, for the adjoint
representation, where $c_{ab}\propto\delta_{ab},$ to which we will
presently specialise. Each such independent invariant can be associated with a
different non-minimal coupling constant, which we absorb into $c_{ab}.$

$\Omega$ is of course real.  Apart
from the $c_{ab}$ term, this form of $K$ is  precisely that to be found
in Wess and Bagger~\cite{Wess:1992cp}, for  the potential in the
Einstein frame, except that in that reference $K$ is defined as follows:
\beq
K=-3M_P^2\log[-\Omega(\phi^a,\phi_b^*)/3].
\label{eq:kformwb}
\eeq
Of course for $c_{ab} =0$ and $|\phi|^2 < 3 M_P^2$, $\Omega < 0$ and 
\eqns{eq:kform}{eq:kformwb}\ are equivalent; we have introduced 
\eqn{eq:kform}\ because we will encounter cases when $\Omega > 0$.

It was shown in~\reference{Einhorn:2009bh}\ that
inclusion of \eqn{eq:bsterm}\  (in the Jordan frame) leads to the form of
$K$ (in the Einstein frame) given in \eqn{eq:kform}.

Now the BS paradigm is to have $\xi \gg 1$, so that the $\xi$ term 
is generally larger than the $\phi^*\phi$ term; it is then important that 
$\Omega$ does not change sign during the inflationary era. In 
\reference{Einhorn:2009bh}\ we presented an example where having $V > 0$ 
during inflation was incompatible with this requirement.

Reverting again to setting $M_P \equiv 1$, the \kahler metric is 
\beq
g_a{}^b = \frac{\pa^2 K}{\pa\phi^a\pa\phi^*_b} = -\frac{3}{\Omega}\delta_a{}^b 
+\frac{3}{\Omega^2}\Omega_a \Omega^b = -\frac{3}{\Omega}\delta_a{}^b 
+ \frac{3}{\Omega^2}(\phi^*_a-\xi c_{ac}\phi^c)(\phi^{b}-\xi c^{bc}\phi^*_c)
\eeq
where
\beq
\Omega_a = \frac{\pa\Omega}{\pa\phi^a}, \quad \Omega^a = \frac{\pa\Omega}{\pa\phi^*_a} 
\eeq
and $c^{ab} = (c_{ab})^*$.
The inverse of the metric is (see Appendix~\ref{inverseK})
\bea
(g^{-1})_a{}^b &=& -\frac{\Omega}{3}\Big(\delta_a{}^b
-\frac{\Omega_a \Omega^b}{\Omega D}\Big)\nonumber\\
&=& -\frac{\Omega}{3}\Big(
\delta_a{}^b- 
\frac{(\phi^*_a -\xi c_{ac}\phi^c)(\phi^b-\xi c^{bc}\phi^*_c)}{\Omega\: D}\Big)\
\eea
where
\beq
\Omega D = \Omega^a\Omega_a -\Omega
= \xi^2 c_{ab}\phi^b c^{ac}\phi^*_c 
- \frac{\xi}{2}(c_{ab}\phi^a\phi^b+c.c.)+3.
\eeq

Substituting our choice of $K$ in \eqn{eq:effpot} we obtain
the following surprisingly simple formula:
\beq
V_F = \frac{9}{\Omega^2}\left[\left|\frac{\pa W}{\pa\phi^a}\right|^2
-\frac{1}{\Omega D} \left|\Omega^a\frac{\pa W}{\pa\phi^a}-3W\right|^2\right]
\label{eq:vfgen}
\eeq
where $\Omega^a = \phi^a - \xi c^{ab} \phi_b^*$.

Obviously interesting is the special class of trajectories such that   
\beq
\Omega^a\frac{\pa W}{\pa\phi^a} = 3W
\label{eq:sp}
\eeq
when $V_F$ is positively semi-definite. The  
minima of the potential on such a trajectory all correspond to zero cosmological constant, 
$V=0$, if there exist solutions to the equations 
\beq
\frac{\pa W}{\pa\phi^a} = 0.
\eeq
These do not necessarily correspond to local minima of the full 
potential $V_F$, however, since the second term in 
\eqn{eq:vfgen}\ may well be negative in the neighbourhood of a solution to 
\eqn{eq:sp}. Other than in the global \sy\ case they also break \sy\ (unless 
$W=0$). 

\section{Scale invariant superpotentials}

In the case when $W$ is a purely cubic superpotential, $V_F$ 
simplifies even more remarkably 
to the following form:
\beq
V_F = 
\frac{9}{\Omega^2}\left[\left|\frac{\pa W}{\pa\phi^a}\right|^2 
- \frac{|\Delta|^2}{\Omega D}\right]\equiv  \frac{9}{\Omega^2}(V_1 + V_2) ,
\label{eq:vsimples}\eeq
where
\beq
\Delta = \xi\frac{\pa W}{\pa\phi^a}c^{ab}\phi^*_b,\qquad 
\Omega D = \Omega^a\Omega_a -\Omega,
\label{eq:deltadef}
\eeq
and we have defined $V_1$ and $V_2$ for later convenience.

\subsection{The $\xi = 0$ case}\label{sec:xizero}

The result \eqn{eq:vsimples} 
is interesting even in the case of minimal coupling when $\xi = 0$.
Then $V_F$ is manifestly positive, with extrema corresponding to $V_F = 0$ 
if there exist solutions to
\beq
\frac{\pa W}{\pa\phi^a} = 0
\label{eq:fterm}
\eeq
such that $\Omega\neq 0$. 

There are three cases to consider.

\begin{itemize}

\item Global \sy, i.e. 
$g_a{}^b = \delta_a{}^b$ in \eqn{eq:scalarL}, and
\beq
V_F = \left|\frac{\pa W}{\pa\phi^a}\right|^2.
\eeq
In this case a solution to \eqn{eq:fterm}\ corresponds to a supersymmetric 
ground state, assuming the $D$-term also vanishes. 
(We review in Appendix~\ref{app:flatness} the fact that an $F$-flat potential 
is also in general $D$-flat, 
unless $W = 0$.) 
Moreover it is easily seen that these correspond 
to the only possible extrema of $V_F$; that is, there are no non-\sic\  
extrema. This is because such extrema would satisfy 
\beq
\frac{\pa^2 W}{\pa\phi^a\pa\phi^b}\frac{\pa W^*}{\pa\phi^*_b} = 0.
\eeq
Now this is a set of homogeneous polynomial equations with each  and
every term of the form $\phi(\phi^*)^2$. Consequently if  this system
has a solution $\phi = \phi_0$, then it will also have a solution  
$\phi = \lambda \phi_0$. But if $\phi \to \lambda\phi$, then $V_F \to 
\lambda^4 V_F$. So the only possible extrema have $V_F = 0$.  

\item Normal supergravity with $\Omega = \phi^*\phi -3$.

Now an extremum satisfying \eqn{eq:fterm} with $\Omega\neq 0$ again has 
$V_F = 0$ but now corresponds to {\it broken\/} \sy, unless also $W =
0$.  In this case (whether or not there are such extrema) there may
exist  extrema with $V \neq 0$ because the extremal condition is no
longer  homogeneous. Obviously, without a $\xi$ term, any new extrema will  have
$\phi \sim M_P$ and so higher order terms in the potentials  become
significant, so it is not clear what physics we can extract from this
case.  

\item Scale invariant supergravity with  $\Omega = \phi^*\phi$.

Once again  an extremum satisfying \eqn{eq:fterm} with $\Omega\neq 0$
again has $V_F = 0$ and  corresponds to {\it broken\/} \sy, unless also
$W = 0$.  Now, however, there may also be $V_F \neq 0$ extrema, because 
the potential is invariant under the rescaling $\phi \to \lambda \phi$.
Of course since in this case we have neither an Einstein 
$M_P^2 R$ term nor a $\xi$ term, we are now describing 
a theory in a background gravitational field.  

\end{itemize}

As in the non-\sic\ case, let us consider the case of a single adjoint
representation of $SU(N)$. The most general cubic superpotential is 
\beq
W = \frac{\sqrt{2}}{3}\lambda\Tr \Phi^3 
=  \frac{1}{3}\lambda d^{ABC}\phi^A\phi^B\phi^C
\label{eq:adjoint}
\eeq
where we can choose $\lambda$ to be real and positive, 
and our normalisation of $d^{ABC}$
is the conventional one for $SU(N)$; see Appendix~\ref{group}.
The crucial difference from the non-\sic\ case is that $\phi^A$ are 
complex fields and correspondingly $\Phi$ is not hermitian (although still 
traceless), and therefore cannot be made real and diagonal 
by a gauge transformation. However, if we assume that the extrema
occur for $V_D=0,$ then 
\beq
\left[ \Phi, \Phi^{\dagger} \right] = 0.
\label{com}
\eeq
Then any $\Phi$ satisfying \eqn{com}\ can be diagonalised 
by a unitary transformation; so we may seek solutions of  
of the form 
\beq
\Phi = \hbox{diag}\, (z_1\cdots z_i \cdots z_N) \quad\hbox{and traceless},
\label{phitrace}
\eeq 
but we cannot in the \sic\ case assume that $z_i$ are real, without 
loss of generality.

Returning to the three cases described in section~\ref{sec:xizero}\ 
we have in turn:

\begin{itemize}

\item Global \sy. 

\eqn{eq:fterm} gives

\beq 
\Phi^2 - \frac{1}{N}T_2 = 0
\label{eq:flatmin}
\eeq

then we find that $z_i$ satisfy the equation
\beq
z^2 = \frac{1}{N}T_2
\eeq
where $T_2 = \sum_i z_i^2$. Since this is a quadratic it has at 
most two solutions, 
$z_i = \pm \sqrt{\frac{T_2}{N}}$, so we see that if $N$ is odd 
there is a unique solution $z_i =0$ 
for all $i$, corresponding to a \sic\ ground state.
If $N$ is even, one has an  additional \sic\ ground state   
\beq
\Phi = \Lambda\, \hbox{diag}\,  (1,-1,\cdots 1,-1),
\label{eq:Neven}
\eeq
(where $\Lambda$ is complex in general),
corresponding to the breaking 
$SU(N) \to SU(\frac{N}{2})
\otimes SU(\frac{N}{2})$.

\item Normal supergravity with $\Omega = \phi^*\phi -3$.

In this case the scalar potential is
\beq
V_F = \frac{18\lambda^2}
{\Omega^2}\left[\Ttil_4-\frac{1}{N}|T_2|^2\right]
\label{vnoxi}
\eeq
where $\Ttil_4 = \Tr \Phi^2 \Phi^{\dagger 2},$ $\Omega=\Ttil_2 - 3,$ with
$\Ttil_2 = \Tr \Phi  \Phi^{\dagger}$.
If we again seek a solution of the form of \eqn{phitrace} then we find 
\beq
z z^{* 2}-\frac{1}{N}T_2^* z      
-\frac{1}{\Omega}z^*
\left(\Ttil_4 - \frac{1}{N}|T_2|^2\right) -\frac{1}{N}\Ttil_3^* = 0
\label{eq:complexcase}
\eeq
where $\Ttil_3^* = \Tr \Phi \Phi^{\dagger 2}$. 
Even for low values of $N,$ it is difficult to analyse the general complex 
solutions of this equation.
For real $z \to x$, \eqn{eq:complexcase}\ reduces to 
\beq
x^3-\frac{1}{T_2-3}\left(T_4 - 3\frac{T_2}{N}\right)x-\frac{1}{N}T_3 = 0.
\label{eq:sugracase}
\eeq
It is easy to show that for even $N,$ \eqn{eq:sugracase} is satisfied 
by \eqn{eq:Neven}. Moreover it follows in general from 
\eqn{eq:sugracase} that $T_4 = T_2^2/N,$ so there are no 
other non-trivial solutions of \eqn{eq:sugracase}. Therefore, for odd $N$, 
there is a unique \sic\ ground state with $\Phi = V_F = 0$; while for 
even $N$ there is an additional \sic\ extremum satisfying 
\eqn{eq:Neven} and  
$V_F = 0$, breaking $SU(N) \to SU(\frac{N}{2})
\otimes SU(\frac{N}{2})$. This extremum is also \sic\ since $W=0$.

\item Scale invariant supergravity with  $\Omega = \phi^*\phi$.

In this case the scalar potential is
\beq
V_F = \frac{18\lambda^2}
{\Ttil_2^2}\left[\Ttil_4-\frac{1}{N}|T_2|^2\right]
\label{vnoxib}
\eeq
leading to \beq
z z^{* 2}-\frac{1}{N}T_2^* z       
-\frac{1}{\Ttil_2}z^*\left(\Ttil_4 - \frac{1}{N}|T_2|^2\right)
-\frac{1}{N}\Ttil_3^* = 0.
\label{eq:sugracasec}
\eeq
In this case $V_F$ is not well defined at $\Phi = 0.$
As usual, it is easier to examine this equation in the special case of 
real $z \to x$:
\beq
x^3 - \frac{T_4}{T_2}x - \frac{1}{N}T_3 = 0.
\label{extnoscale}
\eeq
Since this is a cubic there are at most 
three distinct solutions for $x_i$. 
For even $N$ we again have the \sic\ extremum 
\eqn{eq:Neven}, with $V_F = 0$.

A general classification of even the real extrema is more complicated in
this case.  Let us consider $SU(5)$.
Suppose one of the $x_i$ is zero. It follows from \eqn{extnoscale}\ 
that 
$T_3 = 0$. It is therefore easy to see that the possible solutions 
for $SU(5)$ are $\Phi = \hbox{diag}\,  (1,-1,0,0,0)$, 
$\Phi = \hbox{diag}\,  (1,-1,1,-1,0)$.
with results for $V$ of $V_F^{(1)} = 27\lambda^2/5$ and 
$V_F^{(1)} = 9\lambda^2/10$ respectively. 

If none of the $x_i$ are zero then we have the possible forms 
\bea
A: \Phi &=& \hbox{diag}\,  (1,1,1,z,-3-z)\\
B: \Phi &=& \hbox{diag}\,  (1,1,z,z,-2-2z).
\eea
Substitution of these forms in \eqn{extnoscale}\ and plotting $V$ 
as a function of $z$ (or seeking consistent solutions 
for all $x_i$ of \eqn{extnoscale}) reveals that there are in fact 
solutions corresponding to 
\bea
\label{eq:321}
A: \Phi &=& \hbox{diag}\,  (1,1,1,-3/2,-3/2)\\
B: \Phi &=& \hbox{diag}\,  (1,1,1,1,-4)
\eea
with $V_A = 3\lambda^2/5$ and $V_B = 81\lambda^2/10$ respectively.

Solution~A from~\eqn{eq:321}, corresponding to breaking to 
$SU(3)\otimes SU(2) \otimes U(1),$  has the lowest energy.  
In Appendix~\ref{sec:stability}, we show that this is in fact
stable against quadratic fluctuations.

\end{itemize}

\subsection{The $\xi \neq 0$ case}

The $\xi \neq 0$ case is different because $V_F$ is no longer positive
definite.  However, we show in Appendix~\ref{app:positivity} that 
the leading term in $\xi$ is nonnegative in scale invariant models.

Let us return to the single adjoint case, \eqn{eq:adjoint}.
It is interesting that the form of the results are quite similar in some 
cases to the $\xi = 0$ case.  

For real fields $\phi$, the potential becomes (for $SU(N)$)
\beq
V_F = \frac{18\lambda^2}{\left[(1-\xi)T_2-3\right]^2}\left[T_4-\frac{1}{N}T_2^2
-\frac{\xi^2T_3^2}{\xi(\xi-1)T_2+3}\right].
\label{vrealsimple}
\eeq
In the large $\xi$ limit then, for real $\phi$, 
$V_F$ becomes
\beq
V_F = \frac{18\lambda^2}{\xi^2 T_2^2}
\left[T_4 -\frac{1}{N}T_2^2 
-\frac{T_3^2}{T_2}\right].  
\label{eq:bigxi}
\eeq
Once again we assume that vanishing of $V_D$ implies \eqn{com}, and 
hence a diagonal $\Phi$, 
so we seek an extremum of $V_F$ with $\Phi$ 
of the form 
\beq
\Phi = \hbox{diag}\,  (x_1\cdots x_i \cdots x_N) \quad\hbox{and traceless}.
\label{phitraceb}
\eeq 
We find that $x_i$ satisfy the cubic equation
\beq
2 x^3 T_2^2- 3 T_3 T_2  x^2 + (3 T_3^2-2 T_4 T_2) x
+\frac{1}{N}T_3 T_2^2 = 0.
\label{extremum}
\eeq
Since this is a cubic there are at most 
three distinct solutions for $x_i$. 

Suppose one of the $x_i$ is zero. It follows from \eqn{extremum}\ that 
$T_3 = 0$. It is therefore easy to see that the potential solutions 
for $SU(5)$ are $\Phi = \hbox{diag}\,  (1,-1,0,0,0)$, 
$\Phi = \hbox{diag}\,  (1,-1,1,-1,0)$.
with results for $V_F$ of $V_F = 27\lambda^2/(5\xi^2)$ and 
$V_F = 9\lambda^2/(10\xi^2)$ respectively. 

In both the above cases the 
potential is unstable against quadratic fluctuations. 
One can see this simply by substituting, for example,  
$\Phi = \hbox{diag}\,  (1+Y,-1+Y,-2Y,0,0)$
in the first case, or $\Phi = \hbox{diag}\,  (1,-1,1,-1-X,X)$
in the second. 

If none of the $x_i$ are zero then we have the possible forms 
\bea
A: \Phi &=& \hbox{diag}\,  (1,1,1,z,-3-z)\\
B: \Phi &=& \hbox{diag}\,  (1,1,z,z,-2-2z).
\eea
Substitution of these forms in \eqn{extremum}\ and plotting $V$ 
as a function of $z$ (or seeking consistent solutions 
for all $x_i$ of \eqn{extremum}) reveals that there are in fact 
solutions corresponding to 
\bea
A: \Phi &=& \hbox{diag}\,  (1,1,1,-3/2,-3/2)\\
B: \Phi &=& \hbox{diag}\,  (1,1,1,1,-4)
\eea
with in both cases $V_F^{(1)} = 0$.

Once again, the $SU(3)\otimes SU(2) \otimes U(1)$ trajectory is stable against 
quadratic fluctuations. For example, if we set
\beq
\Phi = \hbox{diag}\,  (
2A+X+Y, 2A -X+Y, 2A+Y,-3A+Y,-3A-4Y)
\eeq
and expand the full potential as a power series in $(X,Y)$ we get 
\bea
V_F &=& \lambda^2\frac{18 (-450A^4\delta +4500A^6
) \delta^3}
{5(30 A^2  \delta - 30 A^2  - 3 \delta)^2 
(30 A^2\delta - 30 A^2  - 3 \delta^2)}\nonumber\\
&+&\lambda^2\delta^2(\frac{2}{A^2}X^2 + \frac{25}{2A^2}Y^2) + \cdots 
\eea
where here $\delta = 1/\xi$, and dropping terms of $O(\delta^3)$ 
from the $X^2$, $Y^2$ terms.
For $\xi A^2 \gg 1$ this becomes
\beq
V_F \sim -\frac{3}{5\xi^3}\lambda^2 +\frac{\lambda^2}{\xi^2}(\frac{2}{A^2}X^2 
+ \frac{25}{2A^2}Y^2).
\eeq
It appears that the leading $1/\xi^3$ term comes entirely from 
the $|W|^2$ term in $V$, as its sign flips if we flip the sign 
of this term. 
Obviously, however,  
making $\xi$ negative corresponds (for large $\xi A^2$)
to $\Omega > 0$ and hence we will encounter a zero in $\Omega$ as 
$A$ decreases
which is clearly problematic, unless we
abandon matching the large $\phi$ and small $\phi$ regions. 

Notice that the result for $V~\sim M_P^4/\xi^3$ is $O(M_{\rm \gut}^4)$ 
if $\xi \sim 10^4$. So perhaps we could have a (larger) positive 
cosmological constant of  $O(M_{\rm \gut}^4)$, which would allow us to
have inflation  and have that cancelled by the $SU(5)$ breaking? But
apart from the fine-tuning  issue, if the $SU(5)$ breaking preserves
\sy\ then it would generate a cosmological  constant that was naturally
of $O(M_{\rm \gut}^6/M_P^2)$.

\section{Scale Invariant Case}

It is interesting to consider the scale invariant case, corresponding in the Jordan frame 
to the absence of the Ricci scalar term, $M_P^2 R$.

We can recover this from the more general results by 
simply replacing $\Omega$ by
\beq
\Omega= \frac{1}{M_P^2}\left[{\textstyle\sum}\phi_a^*\phi^a
-\half\xi(c_{ab}\phi^a\phi^b+c.c.) 
\right]
\eeq
and hence $\Omega D$ by 
\beq
\Omega D = \xi^2 c_{ab}\phi^b c^{ac}\phi^*_c 
- \frac{\xi}{2}(c_{ab}\phi^a\phi^b+c.c.).
\eeq
We find from \eqn{eq:effpot} that (for a scale 
invariant, {\it i.e.,} cubic, superpotential), the scalar potential 
still satisfies \eqns{eq:vsimples}{eq:deltadef}.

\subsection{The adjoint case}

We return to the single adjoint case, that is the potential 
in \eqn{eq:adjoint}.  We find from \eqn{eq:vsimples}
\beq\label{eq:vscale}
V_F = \frac{18\lambda^2}{\Omega^2}
\left[\Ttil_4 -\frac{1}{N}|T_2|^2  -\frac{\xi|\Ttil_3|^2}{\xi\Ttil_2-\Re(T_2)}\right]
\eeq 
where $\Ttil_4\equiv\Tr[\Phi^\dagger{}^2\Phi^2],$ $\Ttil_3\equiv\Tr[\Phi^\dagger\Phi^2].$
It is easy to check that in the large $\xi$ limit \eqn{eq:vscale} reduces 
 to \eqn{eq:bigxi}; the large $\xi$ limit is the same for the scale invariant case. 

If we again specialise to the case of a real-valued extrema of $V_F$ with $\Phi$ 
of the form of \eqn{phitrace}, 
then, so long as $T_2 \neq 0$, we find that $x_i$ satisfy the cubic equation
\beq
2 x^3 T_2^2- 3\rho T_3 T_2  x^2 + (3 \rho T_3^2-2 T_4 T_2) x
+\frac{3\rho-2}{N}T_3 T_2^2 = 0
\label{extremumsi}
\eeq
where we defined $\rho\equiv {\xi}/({\xi-1})>1.$
It is also interesting to ask what the condition for a supersymmetric 
state is. Again assuming \eqn{phitrace}, we find 
that in supersymmetric states the $x_i$ satisfy the quadratic equation
\beq
x^2-\frac{T_3}{T_2}x - \frac{1}{N}T_2 = 0
\label{quad}
\eeq
with non-vanishing diagonal $N \times N$ solutions of the form 
\bea
\Phi &\sim& (1,-1), \nonumber\\
&\phantom{\sim}& (1,1,-2),\nonumber\\ 
&\phantom{\sim}& (1,1,-1,-1), (1,1,1,-3),\nonumber\\ 
&\phantom{\sim}& (2,2,2,-3,-3),(1,1,1,1,-4) \cdots  
\eea
(This is reminiscent of 
the flat space case for a massive adjoint:
\beq
W = \frac{1}{2}m T_2 + \frac{\lambda}{3}T_3
\eeq  
when the $x_i$ also satisfy a quadratic
\beq
\lambda x^2+ m x - \frac{\lambda}{N}T_2 = 0,
\eeq
but of course in that case the solutions are not scale invariant, and 
$\Phi = 0$ is also a solution.) 

In the $SU(5) \to SU(3)\otimes SU(2) \otimes U(1)$ case, it is easy to show 
from \eqn{eq:vscale}\ that 
\beq
V_F = -\frac{3M_P^4\lambda^2}{5(\xi-1)^3}.
\eeq

Turning to the case of more general (i.e. not necessarily \sic) 
extrema, we require a solution to 
\eqn{extremumsi}. If $T_3 = 0$, then this reduces to 
\beq
x(x^2 T_2 - T_4) = 0
\eeq
with solutions corresponding to 
\bea
\Phi &\sim& (1,-1,0), (1,-1,0,0), (1,-1,0,0,0) \cdots\nonumber\\
&\phantom{\sim}& (1,-1,1,-1,0), (1,-1,1,-1,0,0), 
(1,-1,1,-1,0,0,0)\cdots\nonumber\\
&\phantom{\sim}& (1,-1,1,-1,1,-1,0), (1,-1,1,-1,1,-1,0,0), \cdots
\label{nonsic}
\eea
etc, as well as the \sic\ solutions $\Phi \sim (1,-1)$ etc. identified above. 
The extrema identified in \eqn{nonsic} all correspond to $W = 0$ 
and give a positive $V_F$ 
and \sy\ breaking. 

For $SU(N)$, one can characterise a diagonal form of $\Phi$ 
giving a potential extremum in terms of 3 parameters $m,n,z$ as follows:
\beq
\Phi = \hbox{diag}\,  (\underbrace{1,\cdots, 1}_m,\underbrace{z, \cdots, z,}_n
\underbrace{-\frac{m+nz}{N-m-n}, \cdots -\frac{m+nz}{N-m-n}}_{N-m-n}),
\label{genphi}
\eeq
where there are $m$ entries of $1$ and $n$ entries of $z$.

\subsubsection{The adjoint case for $SU(5)$}

For $N=5$ it is straightforward to list and investigate the various 
possible results for $\Phi$ of the form of \eqn{genphi} which 
correspond to extrema of the potential

\begin{itemize}

\item For the case $N=5$, $m=3$, $n=1$ (or, equivalently, $m=3,$ $n=2,$) we find that, as 
well as \sic\ solutions for $z=1$ and $z=-3/2$, there are 
\sy\ breaking solutions for 
\beq
z= z_{A,A'} = -\frac{3}{2}\pm \half\sqrt\frac{15(8-3\rho)}{27\rho-8},
\label{eq:za}
\eeq
and we assume that $1<\rho<8/3.$

These extrema are not local minima for every direction in field space. 
For example if we set
\beq
\Phi = \hbox{diag}\,  (1,1,1,z_A+X,-3-z_A -X)
\label{eq:Xflucts}
\eeq
Then (for large $\xi$)
\beq
V_F = \frac{25}{18\xi^2} - \frac{6859}{2592\xi^2}X^2 + \cdots,
\eeq
and so these extrema are unstable. 
It is interesting that the solutions \eqn{eq:za}\ 
yields $V_F = 0$ for the specific values 
$\xi = (152 \pm 72\sqrt{6})/125 \approx \pm 2.63$. However this 
\sy-breaking solution is unstable in the same direction as described above
in \eqn{eq:Xflucts}. 

\item  For the case $N=5$, $m=2$, $n=1$ (or, equivalently, $m=2,$ $n=2,$) we find that, 
in addition to \sic\ solutions for $z=1$ and $z=-2/3$, there are
\sy\ breaking solutions as follows:

(a)$z=0$. This is unstable with respect to fluctuations for example if 
\beq
\Phi = \hbox{diag}\,  (1,1,X,-1-X,-1)
\eeq
We find (for large $\xi$)
\beq
V_F = \frac{9}{10\xi^2} - \frac{45}{32\xi^2}X^2 + \cdots
\eeq
(b) 
\beq
z = z_{B,B'} = \frac{9\rho+4\pm\sqrt{5(9\rho+4)(9\rho-4)}}{3(3\rho-2)},
\eeq
where again we take $\rho>1.$
These are unstable with respect to
\beq
\Phi = \hbox{diag}\,  (1,1-X,z_B+ X,-1-\frac{z_B}{2},-1-\frac{z_B}{2})
\eeq
We find (for large $\xi$)
\beq
V_F \approx \frac{25}{9\xi^2} - \frac{0.15}{\xi^2}X^2 + \cdots
\eeq

\item The only possible case not included in the 2 cases above 
(when permutations are included) is $m=1$, $n=1,$ 
 leading to $z=-1.$ (or, equivalently, $m=1,$ $n=3,$ $z=0.$)
This is unstable with
\beq
\Phi = \hbox{diag}\,  (1+X,-1,-X,0,0).  
\eeq
We find (for large $\xi$)
\beq
V_F = \frac{27}{5\xi^2} - \frac{81}{4\xi^2}X^2 + \cdots
\eeq
For general $\xi$ the corresponding expression is 
\beq
V_F = \frac{27}{5(\xi-1)^2} - \frac{81\xi}{4(\xi-1)^3}X^2 + \cdots
\eeq
so that the instability in fact persists for all $\xi > 1$. This 
is in fact true for all the cases described in this subsection. 
\end{itemize}
 
\section{The Witten Model}

In this section we consider  a variation of
the inverted hierarchy
model of Witten~\cite{Witten}-\cite{Yamagishi:1982hy}, 
defined by the superpotential
\bea
W &=& \frac{\la_1}{2}d^{ABC} A^A A^B Y^C 
+  \frac{\la_2}{2}X (A^A A^A - m^2)\nn 
&=&\frac{\lambda_1}{\sqrt{2}} \Tr (A^2 Y) 
+ \frac{\lambda_2}{2} X (\Tr A^2 - m^2)\nn
&=&\labar_1 \Tr (A^2 Y) 
+ \labar_2 X (\Tr A^2 - m^2)
\label{eq:witten}
\eea
where $A,Y$ are $SU(5)$ adjoints and $X$ is a singlet, and 
$\labar_{1,2}$ are the couplings as originally  defined by Witten.  
In its complete 
form, with $m \neq 0$,  \sy{} is broken spontaneously in 
the O'Raifeartaigh manner;
moreover  $SU(5)$ is broken to $SU(3)\otimes SU(2)\otimes U(1)$, with the
scale at which this occurs  being larger than and 
not directly related to $m^2$, generated in fact by
dimensional transmutation 
\cite{Einhorn:1982pp}.

At the minimum of the potential it is straightforward to show that
\beq
A = m \frac{\la_2}{\sqrt{\la_1^2+15\la_2^2}} \, \hbox{diag}\, (2,2,2,-3,-3)
\eeq
and
\beq
Y = \Ybar \, \hbox{diag}\, (2,2,2,-3,-3)
= \frac{\labar_2}{\labar_1}X \, \hbox{diag}\, (2,2,2,-3,-3)
\label{eq:yx}
\eeq
with $X$ undetermined in the tree approximation.

Our variation will be to have $m^2 = 0$ in \eqn{eq:witten},  but with
the $SU(5)$ breaking still generated  in similar fashion\footnote{A
discussion of the $m^2 \to 0$  limit of Witten's model appears in
\reference{Einhorn:1982pp}.} but we can imagine the \sy{} breaking
provided instead  by anomaly mediation, gravity-mediated soft breaking,
or even simply a cosmological fluctuation that gives rise to $A\ne 0.$

For $m^2 = 0$, we find in the notation of \eqn{eq:vsimples} that 
\bea
V_1 &=& \frac{\la_1^2}{2} 
\left[\Tr\left(\{A^\dagger,Y^\dagger\}\{A,Y\}\right) - \frac{4}{5}|\Tr AY|^2 + 
\Tr\left( A^\dagger{}^2A^2\right) - \frac{1}{5} |\Tr A^2|^2\right]\nn
&+& \la_2^2\left[|X|^2\Tr A^\dagger A + \frac{1}{4}|\Tr A^2|^2\right] + 
\sqrt2\la_1 \la_2\Re\left(X^\dagger \Tr A^\dagger\{A,Y\}\right)    
\eea
and 
\beq
\Omega = \sum \Phi^*\Phi -\frac{\xi}{2}\left[ 
\Tr A^2 + c_Y\Tr Y^2 + c_X X^2 + c.c.
\right].
\eeq
We can without loss of generality assume that $\xi$, $c_X$ and $c_Y$ are 
real and positive. In fact, we will require that $\Omega<0$ so that the 
Kahler potential $K=-3\log(|\Omega|/3)$ is not singular.  This assumption implies, in particular, 
that $\xi, c_X\xi, c_Y\xi>1.$
  
For simplicity, we will assume consider only real values of the fields.  
Let us first analyse the potential in the approximation that 
 $A \gg X, Y$. Then 
there is no contribution from $V_2$, and 
\beq
V_F = \frac{9}{\Omega^2}V_1 = \frac{9}{(\xi-1)^2}\left[
\frac{\la_1^2}{2}\left(\frac{T_4}{T_2^2}-\frac{1}{5}\right)+\frac{\la_2^2}{4}
\right].
\eeq
If we seek an extremum of $V_F$ with $A$ of the form 
\beq
A = \hbox{diag}\,  (x_1\cdots x_i \cdots x_5) \quad\hbox{and traceless},
\label{Atrace}
\eeq 
then we find that $x_i$ satisfy the cubic equation
\beq
x^3 T_2- T_4 x
-\frac{1}{5}T_3 T_2= 0.
\label{extremumb}
\eeq
Since this is a cubic there are at most 
three distinct solutions for $x_i$. 
Analysing as before we find that the extrema are 
\beq
(1,-1,0,0,0), (1,-1,1,-1,0),(2,2,2,-3,-3), (1,1,1,1,-4).
\eeq 

The one with the lowest energy is $(2,2,2,-3,-3)$, with 
\beq
V_F = \frac{9}{(\xi-1)^2}\left(\frac{\la_1^2}{60}+\frac{\la_2^2}{4}\right),
\label{eq:vfor321}
\eeq
and it appears to be stable against fluctuations, in $A$ at least. 

Encouraged by this fact, we now turn to the more relevant case that 
$A,X,Y$ are all large and $X,Y$ satisfy \eqn{eq:yx}. 
Moreover we will assume that $A,Y$ are parallel, that is that 
$A = \Abar\, \hbox{diag}\,  (2,2,2,-3,-3)$.
Then we find that (notation from \eqn{eq:vsimples})
\beq
V_1 = \la_1^2 (60\Abar^2\Ybar^2+15\Abar^4)+\la_2^2 (30\Abar^2 X^2 +225 \Abar^4) - 
60 \sqrt{2} \la_1 \la_2 X \Ybar \Abar^2,
\eeq
and substituting from \eqn{eq:yx}, the $\Abar^2 X^2$ 
terms all cancel and we get simply 
\beq
V_1 = 15(\la_1^2 + 15 \la_2^2)\Abar^4.
\eeq

Now we analyse $V_2$ in similar fashion.
From \eqn{eq:deltadef}\ we have that 
\beq
\Delta = 15\xi \left[ 2(\la_2 X - \sqrt{2}\la_1 \Ybar) \Abar^2 
+ (c_X \la_2 X - \sqrt{2}c_Y \la_1 \Ybar) \Abar^2\right],
\eeq
and using \eqn{eq:yx} again we get
\beq 
\Delta = 15 \xi (c_X - c_Y ) \la_2 X \Abar^2.
\eeq
Meanwhile,
\bea
\label{eq:omega}\Omega&=&-30(\xi - 1)\Abar^2-
\left[15\left(c_Y\xi-1\right)\frac{\la_2^2}{\la_1^2}
+ \left(c_X\xi-1\right) \right]X^2\nn
&\equiv&-30\xi(\rho \Abar^2 + \rho'_X X^2),\\
\label{eq:omegad}\Omega D &=& 30\xi(\xi-1)\Abar^2 + 30 c_Y\xi(c_Y\xi-1) \Ybar^2 + c_X\xi(c_X\xi-1)X^2 
\nn
&=&30\xi(\xi-1)\Abar^2+30\left[c_Y\xi(c_Y\xi-1)\frac{\la_2^2}{2\la_1^2}+\frac{1}{30}c_X\xi(c_X\xi-1\right])\nn
&\equiv&  30\xi^2(\rho \Abar^2 + \rho_X X^2).
\eea
For reference, we make the implicit definitions in \eqns{eq:omega}{eq:omegad} explicit:
\bea\label{eq:rhos}
\rho \equiv1 - \frac{1}{\xi},\qquad  \rho'_X \equiv \left(c_Y-\frac{1}{\xi}\right)\frac{\la_2^2}{2\la_1^2}
+ \frac{1}{30}\left(c_X-\frac{1}{\xi}\right),\nn 
\rho_X \equiv c_Y\left(c_Y-\frac{1}{\xi}\right)\frac{\la_2^2}{2\la_1^2}
+\frac{1}{30}c_X\left(c_X-\frac{1}{\xi}\right).  
\eea
Noting that
\beq
\rho_X =  \left(c_Y-\frac{1}{\xi}\right)^2\frac{\la_2^2}{2\la_1^2}+\frac{1}{30}\left(c_X-\frac{1}{\xi}\right)^2
+\frac{\rho'_X}{\xi},
\eeq
it follows that $\rho'_X>0$ implies $\rho_X>0.$

Finally, the contribution of the second term in the potential is
\beq
V_2 = -15\frac{(c_X - c_Y)^2 \la_2^2X^2 \Abar^4}
           {2(\rho \Abar^2 + \rho_X X^2)}.
\eeq
Combining this with the result above for $V_1,$ 
\bea
V_1+V_2&=&15\Abar^4\left[\la_1^2+
\la_2^2\left(15-\frac{(c_X - c_Y)^2 X^2}{2(\rho \Abar^2 + \rho_X X^2)}  
\right)\right]\\
&=&\frac{15\Abar^4}{(\rho \Abar^2 + 
\rho_X X^2)}\left[\left(\la_1^2+15\la_2^2\right)\rho \Abar^2 
+ \left(c_X\la_1^2+15c_Y\la_2^2\right)\rho'_X X^2)   \right].
\eea
The second expression above shows that this is nonnegative for all field values since, as noted earlier, we 
have $\rho, \rho'_X, \rho_X>0.$   This is a rather surprising result.  

We are now ready to analyse the complete potential, 
\bea
V_F &=&  \frac{9}{\Omega^2}(V_1 + V_2)\\
&=& \frac{3\Abar^4\left[\left(\la_1^2+15\la_2^2\right)\rho \Abar^2 + 
\left(c_X\la_1^2+15c_Y\la_2^2\right)\rho'_X X^2) \right]}{20\xi^2(\rho \Abar^2 + 
\rho'_X X^2)^2(\rho \Abar^2 + \rho_X X^2)}\\
&=& \frac{3{\cal{A}}^4\left[\left(\la_1^2+15\la_2^2\right)\rho {\cal{A}}^2 + 
\left(c_X\la_1^2+15c_Y\la_2^2\right)\rho'_X )   \right]}
{20\xi^2(\rho {\cal{A}}^2 + \rho'_X )^2(\rho {\cal{A}}^2 + \rho_X )}
\eea
In the last step, we took advantage of the scale invariance to write 
the result in terms of the ratio of field values ${\cal{A}}=\Abar/X.$  

Although we have $V_F>0$ in general, it is certainly not flat, {\it
i.e.,} not independent of ${\cal{A}}.$ Note that if $X \to \infty$ for
fixed $\Abar$, or ${\cal{A}}\to0,$ then $V_F \to O({\cal{A}}^4)$.   If
we return for a moment to the original model, \eqn{eq:witten}, we recall
that, as $m^2\to0,$ $\Abar\to0,$ but $X$ and $\Ybar,$ remain at their
values determined by dimensional transmutation. Although the model is
supersymmetric in the limit, as remarked earlier, one may simply suppose
that, initially $\Abar\ne 0$ for whatever reason, and work out the
consequences as it relaxes toward its equilibrium value, which has been
assumed to be negligible compared to  the \gut\ scale.  Thus, it is
natural to inquire further into the behaviour of $V_F$ for $\Abar/X\ll1.$
We find
\beq
\left[\frac{135\la_1^4(c_X\la_1^2+15c_Y\la_2^2)^2}{(c_X^2\la_1^2
+15c_Y^2\la_2^2)(c_X\la_1^2+15c_Y\la_2^4)}\right] 
\left(\frac{M_P\Abar}{\sqrt{\xi} X}\right)^4 + 
O\left( \left( \frac{\Abar}{X} \right)^6\right),
\label{eq:flucts}\eeq
up to small corrections of order $1/\xi.$
We have restored the unit of mass, $M_P$ that was introduced in
performing the conformal transformation to Einstein frame.  Since it
serves as the effective Planck mass for the gravitational interaction,
it takes the value of order $\sqrt{\xi} X,$ at least naively, so that 
$V_F$ is indeed correctly represented in units of the gravitational
constant.  It is interesting that the coefficient in square brackets in \eqn{eq:flucts}\ is 
naturally nonnegative for all values of the couplings, which was not
guaranteed {\it a priori,} so far as we are aware\footnote{We have not
investigated whether the curvature remains positive for complex
$\Abar.$}.  This result is therefore like a small field inflationary
model of the form $\lambda \Abar^4.$  For such models to have any chance
at describing a phenomenologically acceptable inflationary epoch, the
coefficient $\lambda$ must be exceedingly small, which could be arranged
but does not seem to be required in the present context. Even if finely
tuned, such models have been rather thoroughly investigated\footnote{For some
discussion and references, see, {\it e.g.,} \reference{BenDayan:2010yz}.} 
and are essentially ruled by the most recent WMAP data.

At the other extreme, for $\Abar >> X$, or ${\cal{A}}\to\infty,$, we obtain
\beq
V_F = \frac{3}{20\xi^2\rho^2} \left[a_1- (a_2+2a_1\rho'_X)\frac{1}{\rho{\cal{A}}^2}\right]
\eeq
where $a_1 = \la_1^2 + 15 \la_2^2$ and $a_2 = (c_X - c_Y)^2 \la_2^2/2$.
So the large $\Abar$-flat trajectory is unstable against fluctuations in
$X$.  On the other hand, although $V_F({\cal{A}})$ approaches its
asymptotic value from below, depending on the values of the various
parameters, it is not obviously monotonically increasing.  This raises
the possibility of a local minimum at some value of the ratio
${\cal{A}}_0,$ which may provide a flat direction of the form
$C(\Abar-{\cal{A}}_0X)^2/X^2,$ where the overall scale $X$ is
undetermined classically but will be determined by dimensional
transmutation in higher order. This then might provide a model for large
field inflation\footnote{Of course, if such a minimum exists, it is
metastable and will eventually tunnel to smaller values of the fields. 
So it would be necessary that its lifetime is long enough to allow
sufficient expansion, typically on the order of 60 e-folds.}.

To explore this possibility, we need to calculate the derivatives of
$V_F$ and determine the range of parameters that generate such a
scenario.  We first seek values of ${\cal{A}}$ where $V_F$ is
stationary.  Since $V_F>0$ everywhere (except at the origin,) we may
equally well consider the variation of $\log(V_F).$  It is convenient to
absorb the factor of $\rho$ into the ratio ${\cal{A}}$ and to take
advantage of the fact the $V_F$ depends on ${\cal{A}}$ only through even
powers.  Defining $w\equiv\rho{\cal{A}}^2$ and 
$a_3\equiv\left(c_X\la_1^2+15c_Y\la_2^2\right)\rho'_X>0,$ 
we may express the potential and its first variation as
\bea
V_F&=&\frac{3}{20\xi^2\rho^2}  \left[\frac{w^2\left(a_1 w + a_3 \right)}{(w + \rho'_X )^2(w+ \rho_X )}\right]\\
\frac{\pa \log[V_F]}{\pa w}&=&\frac{2}{w}+\frac{a_1}{\left(a_1 w + a_3 \right)}
-\frac{2}{(w + \rho'_X )}-\frac{1}{(w + \rho_X )}.
\eea
Thus, we need to solve ${\pa \log[V_F]}/{\pa w}=0.$  At first, one might
think that the  vanishing of the first derivative for $w\ne0,$ requires
the solution of a cubic polynomial;  however, because $V_F$ asymptotes
to a constant, the cubic term is absent, and we only need solve a
quadratic equation.  We find that the first derivative vanishes for 
\beq
\left( a_1\left( 2\rho'_X + \rho_X\right)-a_3 \right)w^2+\rho'_X(3a_1\rho_X+a_3)w+2a_3\rho_X\rho'_X=0.
\eeq
There exist real roots if the discriminant $\widetilde{\Delta}$ is non-negative: 
\beq
\widetilde{\Delta}\equiv\rho'_X \left( a_3-a_1\rho_X\right)
\big( a_3(\rho'_X+8\rho_X)-9a_1\rho_X\rho'_X\big)\ge0,
\eeq
which can be arranged.  
Then the roots take the values
\beq
w_\pm \equiv \frac{-\left(a_3+3a_1\rho_X\right)\rho'_X\pm
\sqrt{\widetilde{\Delta}}}{2\left( a_1( 2\rho'_X + \rho_X)-a_3\right) }
\eeq
Naturally, we expect one root to be a maximum and the other to be a
minimum of $V_F.$ However, recalling that $w\equiv\rho{\cal{A}}^2,$ the
minimum must occur at $w>0$ to be acceptable.  Unfortunately, it turns
out that both roots are negative.  To see this, one may first show,
under the assumptions stated previously, that the denominator $\left(
a_1( 2\rho'_X + \rho_X)-a_3\right)$ is positive, so that obviously
$w_{-} < 0$.  In order for $w_{+}>0,$ we would have to have
$\sqrt{\widetilde{\Delta}}>\left(a_3+3a_1\rho_X\right)\rho'_X,$ or
$\widetilde{\Delta}-\left(a_3+3a_1\rho_X\right)^2\rho'_X{}^2>0.$ 
However, this difference turns out to be equal to $8 a_3
\rho_X\rho'_X(a_3-a_1(2\rho'_X + \rho_X))<0.$  Thus, both roots are at
negative $w,$ so that in fact, $V_F$ is monotonically increasing
throughout the region $w>0.$

One could imagine variants on this theme involving several of the fields
in this model, but, since this sort of model appears to be quite
different from the Higgs inflation models we were seeking in this paper,
we leave such speculations for future work.

\section{Conclusions}

We have shown that in a \gut\ with non-minimal scalar coupling to gravity,
an era of Higgs inflation is possible with the relevant Higgs multiplet
being  the one responsible for breaking of the \gut\ symmetry. In
particular, in a  non-\sic\ $SU(5)$ with an adjoint multiplet, the flat
direction with the lowest energy corresponds to $SU(3) \otimes SU(2)
\otimes U(1)$, and is stable against fluctuations away from this
direction. 

Non-\sic\ $SU(5)$ has problems; with gauge unification and with  proton
decay for example. While it might be possible to  construct a viable
model along these lines, we have in this paper also investigated whether
the above result can be achieved in a \sic\ \gut. There has been 
considerable work on such theories; in particular on $SO(10)$; there  in
our opinion,  however, no really compelling  \sic\ \gut\ exists as yet.
In this paper  we have restricted our attention largely to $SU(5)$;  and
in all the cases we have looked at, although there have existed positive
energy flat directions at large field magnitude, they have invariably
been unstable against quadratic fluctuations, in the same
manner~\cite{Ferrara:2010yw} as the original \nmssm\ model described in
\reference{Einhorn:2009bh}. This problem has been approached in the
literature by including higher order terms in the \kahler\ potential, an
approach that we find unattractive and have tried to avoid. As we
indicated above, the result of this ``purist'' philosophy is,
unfortunately, that  we have been unable to find a \sic\ model that has 
both $V_F>0$ in a flat direction and is stable to fluctuations in other
directions. In our opinion this outcome is likely to persist  in
generalisations to other gauge symmetry groups.

It seems to us, therefore, a minimal extension of the BS scenario to a \gut\ is 
more promising in the non-\sic\ case.  
Whether this is worth pursuing further 
may well depend upon whether the current absence of evidence 
for low energy \sy\ at LHC experiments persists. 

\appendix

\section{Group Theory}
\label{group}

We consider a complex scalar multiplet $\phi^a$ transforming according to 
a (in general reducible) complex representation of a unitary group 
(we will presently specialise to $SU(N)$) as follows: 

\beq
\phi^a \to \phi^{a'} = U^a{}_b\phi^b
\eeq

The complex conjugate of $\phi$,  $\phi^*_a = (\phi^a)^*$ transforms 
as
\beq
\phi^*_a \to \phi^{*'}_a = (U^{-1})^b{}_a\phi^*_b
\eeq
and of course $\phi^*_a \phi^a$ is invariant. 
We see that it is helpful to use 
a notation where complex conjugation raises and lowers the index; a familiar 
notation for the fundamental representation of $SU(N)$.

For the generators of the group in the $\phi^a$ representation we use the notation 
$(R^A)^a{}_b$, thus 
\beq
U = e^{i\alpha^A R^A}
\eeq
and 
\beq
[R^A, R^B] = if^{ABC}R^C.
\eeq
When $\phi^a$ is in the adjoint representation then obviously $\phi^a \to \phi^A$, 
and $(R^A)^b{}_c \to -if^{ABC}$. 

For a single adjoint representation of $SU(N)$ we would have 
the invariant tensor $c_{AB} = \delta_{AB}$, 
so in that case it is tempting to define $\phi_A = \delta_{AB}\phi^B$, but 
since we associate raising and lowering indices with complex conjugation 
we will not do this. Thus whereas when we write 
formulae valid for an arbitrary representation, index summations will always 
involve one up and one down index 
(simply because the product of a representation with its  complex conjugate 
always contains a singlet) in the case of the adjoint we also have the invariants 
$\phi^A\phi^A$ and $\phi_A^*\phi_A^*$. 

It is convenient to write an adjoint representation as 
a $N$-dimensional matrix,  
$\Phi = \frac{1}{\sqrt{2}}\lambda^A \phi^A$. 
where $\lambda^A$ are the generators in the fundamental representation, but conventionally 
defined as $\lambda^A = 2 R^A$. Thus 
\beq
[\lambda^A, \lambda^B] = 2if^{ABC}\lambda^C.
\eeq
 
The following relations are valid for $SU(N)$:
\bea
\Tr \Phi^2 &=& \phi^A \phi^A\\
\Tr \Phi^3 &=& \frac{1}{\sqrt{2}}d^{ABC}\phi^A\phi^B\phi^C\\
\Tr \Phi^4 &=& \frac{1}{2}d^{ACD} \phi^C\phi^D d^{ADE} \phi^D\phi^E 
+ \frac{1}{N} (\phi^A \phi^A)^2.
\eea
When we generalise to complex $\phi$ we will also need
\bea
\Tr \Phi\Phi^{\dagger} &=& \phi^A \phi_A^*\\
\Tr \Phi^2\Phi^{\dagger} &=& \frac{1}{\sqrt{2}}d^{ABC}\phi^A\phi^B\phi_C^*\\
\Tr \Phi^2\Phi^{\dagger 2} &=& 
\frac{1}{2}d^{ACD} \phi^C\phi^D d^{ADE} \phi_D^*\phi_E^* 
+ \frac{1}{N} |\phi^A \phi^A|^2
\eea
All these expressions follow easily from the 
formula
\beq
\{\lambda^A, \lambda^B\} = 2d^{ABC}\lambda^C + \frac{4}{N}\delta^{AB}.
\eeq

\section{Inverse Metrics}
\label{inverseK}

The fact that the inverse of a \kahler 
metric 
of the form
\beq
g_a{}^b = \delta_a{}^b 
-\frac{1}{\Omega}\Omega_a \Omega^b
\label{eq:kr}
\eeq
is
\beq
(g^{-1})_a{}^b = \delta_a{}^b
-\frac{\Omega_a \Omega^b}{\Omega D}
\label{eq:krinv}\eeq
where 
\beq
D =\sum \frac{|\Omega_a|^2}{\Omega}-1
\eeq
is easily derived using the fact that 
\beq
\Pi_1 \equiv \delta_a{}^b - \frac{\Omega_a \Omega^b}{\sum |\Omega_c|^2} 
\eeq
and 
\beq
\Pi_2 \equiv \frac{\Omega_a \Omega^b}{\sum |\Omega_c|^2}
\eeq
are projection operators,
\beq
(\Pi_1)^2=\Pi_1, (\Pi_2)^2=\Pi_2, \Pi_1 \Pi_2 = 0, \Pi_1 + \Pi_2 = 1.
\eeq
Thus 
\bea
g_a{}^b &=& \delta_a{}^b - \frac{\Omega_a \Omega^b}{\sum |\Omega_c|^2}
+\frac{\Omega_a \Omega^b}{\sum |\Omega_c|^2}
-\frac{\Omega_a\Omega^b}{\Omega}\nonumber\\
&=&
\Pi_1 +\left(\frac{\Omega-\sum |\Omega_c|^2}{\Omega}\right)\Pi_2
\label{eq:krpi}
\eea
whence \eqn{eq:krinv}\ follows, using 
\beq
\left(a_1\Pi_1 + a_2 \Pi_2\right)^{-1} = \frac{1}{a_1}\Pi_1 + \frac{1}{a_2}\Pi_2.
\eeq 
It is easy to verify that 
\beq
g_a{}^b (g^{-1})_b{}^c = (g^{-1})_a{}^b g_b{}^c = \delta_a{}^c.
\eeq 

It is interesting (if not immediately relevant to our considerations here) 
to generalise the above case to the problem of finding the inverse 
of a matrix whose components are either the identity $\delta^{ab}$ or the outer product 
of the vector $\{A^a, A^{*a} \}$ with itself.

For this discussion raising and lowering indices by complex conjugation 
is no longer convenient. Consider the matrix  
\beq
g^{ab} = \sum_i a_i P_i
\eeq
where 
\beq
P_1 = \delta^{ab}, P_2 = A^a A^{*b}, P_3 = A^{*a} A^b,
P_4 =  A^a  A^b, P_5 = A^{*a}A^{*b}.
\eeq
The coefficients $a_i$ may depend on scalar invariants involving this vector,
such as $\Omega.$  The matrix $g^{ab}$ is hermitian if $a_{1,2,3}$ are real and $a_5 = a_4^*$. 

It is easy to construct a multiplication table for the $P_i$:
\begin{table}
\begin{center}
\begin{tabular}{|c| c c c c c|} \hline
      & $P_1$ & $P_2$        & $P_3$       & $P_4$ & $P_5$\\ \hline
$P_1$ & $P_1$ & $P_2$        & $P_3$       & $P_4$ & $P_5$ \\ \hline
$P_2$ & $P_2$ & $\alpha P_2$ & $\beta P_4$ & $\alpha P_4$ & $\beta P_2$ \\ \hline
$P_3$ & $P_3$ & $\gamma P_5$ & $\alpha P_3$& $\gamma P_3$ & $\alpha P_5$ \\ \hline
$P_4$ & $P_4$ & $\gamma P_2$ & $\alpha P_4$  & $\gamma P_4$ & $\alpha P_2$ \\ \hline
$P_5$ & $P_5$ & $\alpha P_5$ & $\beta P_3$  & $\alpha P_3$ & $\beta P_5$ \\ \hline
\end{tabular}
\end{center}
\caption{\label{anomfreeb}Multiplication Table}
\end{table}
In Table~\ref{anomfreeb}, $\alpha = A^{*a}A^a$, $\beta = A^{*a}A^{*a}$,
$\gamma = A^a A^a$, and the $(i,j)$ element of the array is $P_i P_j$.

Armed with this table it is straightforward to construct the inverse 
of $g^{ab}$, it is 
\beq
(g^{-1})^{ab} = \sum_i b_i P_i
\eeq
where 
\bea
b_1 &=& \frac{1}{a_1}\nonumber\\
b_2 &=& -\frac{a_1 a_2+ \alpha a_2 a_3 - \alpha a_4 a_5}{\Delta}\nonumber\\
b_3 &=& -\frac{a_1 a_3 + \alpha a_2 a_3 - \alpha a_4 a_5}{\Delta}\nonumber\\
b_4 &=& -\frac{a_1 a_4 - \beta a_2 a_3 + \beta a_4 a_5}{\Delta}\nonumber\\
b_5 &=& -\frac{a_1 a_5 - \gamma a_2 a_3 - \gamma a_4 a_5}{\Delta}
\eea
and
\beq
\Delta = a_1 [ a_1^2 + \alpha a_1 a_2 + \gamma a_1 a_4 - \beta \gamma a_2 a_3 
+\beta a_1 a_5 +\beta\gamma a_4 a_5 + \alpha^2 a_2 a_3 
+ \alpha a_1 a_3 -\alpha^2 a_4 a_5]
\eeq 
The inverse \eqn{eq:krinv} of the metric \eqn{eq:kr}\ is easily derived, 
by setting 
$a_1 = 1, a_2 = -1/\Omega$, $a_{3,4,5} = 0$.

\section{$F$ and $D$-flatness}
\label{app:flatness}

The conditions for an unbroken \sic\ state are 
\beq
F_i = D^a = 0
\eeq
where on a curved background,
\beq
F_i = W_i + K_i W,
\eeq
and 
\beq
D^a = G_i (R^a)^i{}_j\phi^j 
\eeq
where 
\beq
G = K + \ln W + \ln W^* 
\eeq
We see that 
\beq
G_i = (1/W)F_i
\eeq
so that unless $W = 0$, $F$-flatness implies $D$-flatness.

On flat space the argument is more tricky.
Here we have 
\beq
F_i = W_i 
\eeq
and 
\beq
D^a = \phi^*_i (R^a)^i{}_j\phi^j 
\eeq
and vanishing of $F_i$ does not seem to tell us much about $D^a$.
 
However, consider the gauge transformation $\phi \to \phi' = U\phi$.
We have $W(\phi') = W(\phi)$ and hence $(F_i)' = 0$ if $F_i = 0$. 
But since $W$ is holomorphic we can transform 
$\phi^* \to \phi^{*'} = V\phi^*$ for $V \neq U$.

Consequently 
\beq
D^a \to \phi^{\dagger}V^{\dagger} (R^a) U\phi
\eeq
and $U$, $V$ can be chosen so that $D^a$ transforms to zero. 
This is easy to see; given a gauge invariant polynomial 
$P(\phi)$ it is trivial that 
\beq
\frac{\pa P}{\pa \phi} R \phi = 0,
\eeq
and so we simply have to  choose $\frac{\pa P}{\pa \phi}$ so that 
\beq
\frac{\pa P}{\pa \phi} = \phi^{\dagger}V^{\dagger}
\eeq

\section{Stability of the Scale-Invariant, $SU(5)$ Minimum}
\label{sec:stability}

In this appendix, we show that the lowest energy extremum, Solution~A
from~\eqn{eq:321}, is in fact a local minimum. In order to enforce the
trace constraint, $\textstyle{\sum} z_j=0,$ the minimum of the scalar
potential in \eqn{vnoxib} may be obtained by the method of Lagrange
multipliers.  We seek to find extrema of the auxiliary function $G,$
\beq
G(z,z^*)=V_F(z,z^*)-C^*\sum z_j -C\sum z_j^*.
\eeq
As usual, when minimising $G,$ we ignore the constraint condition on the variables, 
and then choose the (complex) constant $C$ so as to enforce the trace constraint.  Thus,
\bea
\frac{\pa G}{\pa z_k}&=&\frac{\pa V_F}{\pa z_k}-C^*=0,\\
\frac{\pa V_F}{\pa z_k}&=&\frac{18\lambda^2}{\Ttil_2^2}\left[ z_k z_k^{* 2}-\frac{1}{N}T_2^* z_k       
-\frac{1}{\Ttil_2}z_k^*\left(\Ttil_4 - \frac{1}{N}|T_2|^2\right)
 \right].
\eea
If one sums the first equation over $k$, enforcing the trace condition, we find
\beq
C^*=\frac{18\lambda^2}{\Ttil_2^2 N}\Ttil_3^*,
\eeq
resulting in the root equation \eqn{eq:sugracasec}.

One may use the auxiliary function $G$ to explore the second variation as well.  
Suppose we expand about an extremum $\hat{z}$ of the form of \eqn{phitrace}. 
Writing $z=\hat{z}+\delta z,$ then it can be shown that 
\beq\label{eq:vsecond}
\delta^{2}G=\delta^{2}V_F=\half\sum_{i,j} \left[\frac{\pa^2 G}{\pa z_i \pa z_j}\Big|_{\hat{z}}\delta z_i \delta z_j + 
\half\frac{\pa^2 G}{\pa z_i \pa z_j^*}\Big|_{\hat{z}} \delta z_i \delta z_j^*\right] + c.c.,
\eeq
where the partial derivatives are at $\hat{z}, \hat{z}^*,$ and the
variations obey the constraint $\textstyle{\sum}\delta z_k=0.$  In other
words, one may treat the components $z_k$ as independent and $C$ as a
constant in carrying out the derivatives, provided one enforces the
constraint condition at the end\footnote{This statement is true for
nonlinear constraints as well.}.  Therefore, we simply need to calculate
the matrix of second derivatives (or Hessian) of $V_F:$
\bea
\label{eq:hessian1}
\frac{\pa^2 G}{\pa z_i^* \pa z_j^*}&=&\frac {36\lambda^2}{N\Omega^2}
\left[ \delta_{ij}\left(Nz_j^2-T_2\right)-a^2 z_i z_j-2\frac{\widetilde{T_3}}{\Omega}\left(z_i+z_j\right)\right], \\
\label{eq:hessian2}
\frac{\pa^2 G}{\pa z_i \pa z_j^*}&=&\frac {36\lambda^2}{N\Omega^2}
\left[ \delta_{ij}\left(2N|z_j|^2-a^2\Omega\right)-2z_i z_j^*-a^2z_i^* z_j - 
\frac{2}{\Omega}\left(z_j \widetilde{T_3}^*+ z_i^*\widetilde{T_3}\right)\right]\!,
\eea
where 
\beq
a^2 \equiv \kappa^2-\kappa_2^2,\quad \kappa^2\equiv \frac{N \Ttil_4}{\Omega^2} 
\quad \kappa_2\equiv \frac{|T_2|}{\Omega}.
\eeq
One can show that $\kappa^2\ge1\ge\kappa_2,$ so that $a^2\ge0.$
Some remarks are in order in how we arrived at these expressions.  Since
each component $\hat{z}_k$ satisfies the root equation
\eqn{eq:sugracasec}, wherever we encountered a cubic such as
$z_j^2z_j^*,$ we replaced it with the corresponding linear terms using
the root equation in the form
\beq\label{eq:root}
N z^2 z^*= z\, a^2 \Omega+z^*\,T_2+\widetilde{T_3}.
\eeq

It is often convenient to decompose the second variation \eqn{eq:vsecond} 
in terms of real and imaginary parts, $\delta z_i\equiv\delta x_i+i\delta y_i.$  
\bea\label{eq:realsecond}
\delta^{2}V_F&=&\Re\!\left[G_{i\bar{j}}+G_{ij}\right]\delta x_i\delta x_j+
2\Im\! \left[G_{i\bar{j}}-G_{ij}\! \right]\delta x_i\delta y_j
+ \Re\! \left[G_{i\bar{j}}-G_{ij}\! \right]\delta y_i\delta y_j\nn
&\equiv&\half
\begin{pmatrix}\delta x_i & \delta y_i\end{pmatrix}
\begin{pmatrix}
A_{ij}  & C_{ij}\\ 
C_{ji} & B_{ij}
\end{pmatrix}
\begin{pmatrix} \delta x_j \\ \delta y_j\end{pmatrix},
\eea
where each matrix is obviously symmetric and real. 
However, we must recall that these variations are constrained by the traceless condition.  
Perhaps the easiest way to take that into account is to explicitly 
eliminate one of these coordinates, say,
\beq\label{eq:tracelss}
\delta x_1=-\sum_2^N \delta x_\alpha, \quad \delta y_1=-\sum_2^N \delta y_\alpha.
\eeq
  Then the problem reduces to analysing the independent $\!(N\!-\!1)$-dimensional variations
\beq\label{eq:realreduced}
\half
\begin{pmatrix}\delta x_\alpha & \delta y_\alpha\end{pmatrix}
\begin{pmatrix}
\widetilde{A}_{\alpha\beta}  & \widetilde{C}_{\alpha\beta}\\ 
\widetilde{C}_{ji} & \widetilde{B}_{\alpha\beta}
\end{pmatrix}
\begin{pmatrix} \delta x_\beta \\ \delta y_\beta\end{pmatrix},
\eeq
where $\widetilde{A}_{\alpha\beta}\equiv A_{\alpha\beta}
+A_{11}-A_{1\beta}-A_{\alpha1},$ and similarly for 
$\widetilde{B}_{\alpha\beta}$ and $\widetilde{C}_{\alpha\beta}.$
For real extrema, $\widetilde{C}_{\alpha\beta}=0,$ so the second variation 
obviously factors into the sum of separate  variations of the real and imaginary parts. 

Finally, for the case of interest, $\hat{z}$ is real and of the form of \eqn{eq:321}.  
Taking $\hat{x}=\{2,2,2,-3,-3\}$ and $N=5,$  we find 
\bea
A_{ij}&\propto& \left[\frac{5}{3}\delta_{ij}\left(3\hat{x}_j^2-7\right)-
\frac{7}{9}\hat{x}_i \hat{x}_j+\frac{4}{3}(\hat{x}_i+\hat{x}_j)\right],\nonumber \\
B_{ij}&\propto& \left[\frac{5}{3}\delta_{ij}\left(\hat{x}_j^2+5\right)-
\frac{2}{3}\hat{x}_i \hat{x}_j\right].
\eea
One may then easily compute the constrained variation
\eqn{eq:realreduced} by eliminating one of the components of $\delta
x_k$ and $\delta y_k.$  One then finds each matrix
$\widetilde{A}_{\alpha\beta}, \widetilde{B}_{\alpha\beta}$ has one zero
eigenvalue corresponding to displacements proportional to $\hat{x},$
plus 3 positive eigenvalues, showing that fluctuations in directions
other than the flat direction are stable.  

Of course, neglecting loop corrections to the potential, the magnitude
of the field $\Phi$ is completely arbitrary, by classical scale
invariance, so that $\phi \hat{x}$ is a good candidate for an
inflationary field.  Scale invariance will be broken at one-loop, and
one would expect the scale $\phi$ to be determined via dimensional
transmutation.

\section{$V_F$ for Large $\xi$}\label{app:positivity}

Here we show that, in scale invariant models, $V_1+V_2\ge0$ to leading order in 
$\xi$ and that the correction terms are negative.    
Starting from \eqns{eq:vsimples}{eq:deltadef}, we may write 
\beq
V\equiv V_1+V_2=
\frac{1}{\Omega D}\left[\Omega D \left| \frac{\pa W}{\pa\phi^a} \right|^2 -
|\Delta|^2\right] 
\label{eq:xione}
\eeq
with 
\bea \label{eq:Omega}
\Omega &=& \frac{1}{M_P^2}\left[{\textstyle\sum}\phi_a^*\phi^a
-\half\xi(c_{ab}\phi^a\phi^b+c.c.) 
\right]\\
\label{eq:omegaD}
\Omega D &=& \xi^2 c_{ab}\phi^b c^{ac}\phi^*_c 
- \frac{\xi}{2}(c_{ab}\phi^a\phi^b+c.c.)\\
\label{eq:delta}
\Delta &=& \xi\frac{\pa W}{\pa\phi^a}c^{ab}\phi^*_b.
\eea
(Here, $M_P$ is an arbitrary scale.)
Inserting Eqs.~(\ref{eq:omegaD}),(\ref{eq:delta})\ into \eqn{eq:xione}, 
we find for the quantity in brackets
\beq\label{eq:brackets}
\left[\xi^2 \left(\left(c_{ab}\phi^b c^{ac}\phi^*_c\right)  
W_dW^d- \left|W_a c^{ab}\phi^*_b\right|^2\right) - 
\frac{\xi}{2}(c_{ab}\phi^a\phi^b+c.c.)W_dW^d\right],
\eeq
where we have defined 
\beq
W^a = \frac{\pa W^*}{\pa\phi^*_a}.
\eeq
The $O(\xi^2)$ terms are nonnegative by the Cauchy-Schwarz inequality.  Further, 
it vanishes if and only if  $W_a\propto \phi^*_a$ for the extremum values of the fields.  

The factor in front, $1/\Omega D,$ is positive and proportional to $\xi^2$ for large $\xi.$  
Therefore, $V$ approaches a non-negative constant asymptotically:  
\beq \label{eq:vasymp}
V\to  W_dW^d- 
\frac{\left|W_a c^{ab}\phi^*_b\right|^2} {c_{ab}\phi^b c^{ac}\phi^*_c}. \eeq

Concerning the first correction terms of order $1/\xi,$ they come from two sources: 
the second term in \eqn{eq:brackets} and the second term in \eqn{eq:omegaD} 
in the factor in front.  When combined, we find the leading correction to \eqn{eq:vasymp}
\beq
-\frac{1}{\xi}
\frac{\left|W_a c^{ab}\phi^*_b\right|^2} {(c_{ab}\phi^b c^{ac}\phi^*_c)^2}
\Re \left(c_{ab}\phi^a\phi^b\right).
\eeq
 Under our assumptions, the factor  
$\Re \left(c_{ab}\phi^a\phi^b\right)$ 
is positive, so the $O(1/\xi)$ term above is negative. 
However, $V_F=9V/\Omega^2$ in Einstein frame, so all such models have 
$V_F\to O(1/\xi^2)$ as $\xi\to\infty.$  The leading term is 
\beq
V_F\to \frac{9}{\xi^2\left(\Re c_{ab}\phi^a\phi^b\right)^2} 
\left(W_dW^d- 
\frac{\left|W_a c^{ab}\phi^*_b\right|^2}{c_{ab}\phi^b c^{ac}\phi^*_c} \right).
\eeq

\section*{\large Acknowledgements}

This work was partially supported by 
the Science and Technology Research Council
under Grant No. ST/J000493/1 and by 
the National Science Foundation
under Grant No. PHY11-25915. While part of it was done one of us (DRTJ) was 
visiting the Aspen Center for Physics.


\begin{thebibliography}{99}

\bibitem{Bezrukov:2007ep}
  F.~L.~Bezrukov and
M.~Shaposhnikov,
Phys.\ Lett.\  B {\bf 659} (2008) 703 
  [arXiv:0710.3755 [hep-th]].
  

\bibitem{Barvinsky:2008ia}
  A.~O.~Barvinsky, A.~Y.~Kamenshchik and A.~A.~Starobinsky,
  JCAP {\bf 0811}  (2008) 021
  [arXiv:0809.2104 [hep-ph]].  
\bibitem{DeSimone:2008ei}
  A.~De Simone, M.~P.~Hertzberg and F.~Wilczek,
  Phys.\ Lett.\  B {\bf 678}  (2009) 1
  [arXiv:0812.4946 [hep-ph]].
  
\bibitem{Bezrukov:2008ej}
  F.~L.~Bezrukov, A.~Magnin and M.~Shaposhnikov,
  Phys.\ Lett.\  B {\bf 675} (2009) 88
  [arXiv:0812.4950 [hep-ph]].
\bibitem{Burgess:2009ea}
 C.~P.~Burgess, H.~M.~Lee and M.~Trott,
  JHEP {\bf 0909}  (2009) 103   
  [arXiv:0902.4465 [hep-ph]].

\bibitem{Barbon:2009ya}
  J.~L.~F.~Barbon and J.~R.~Espinosa, 
  Phys.\ Rev.\  D {\bf 79} (2009) 081302 
  [arXiv:0903.0355 [hep-ph]].
\bibitem{Bezrukov:2009db}
  F.~Bezrukov and M.~Shaposhnikov,
  JHEP {\bf 0907}  (2009) 089
  [arXiv:0904.1537 [hep-ph]].
  
  
\bibitem{Barvinsky:2009fy}
 A.~O.~Barvinsky, A.~Y.~Kamenshchik, C.~Kiefer, A.~A.~Starobinsky and
C.~Steinwachs,
  JCAP {\bf 0912}  (2009) 003
 [arXiv:0904.1698 [hep-ph]].

\bibitem{Barvinsky:2009ii}
A.~O.~Barvinsky, A.~Y.~Kamenshchik, C.~Kiefer, A.~A.~Starobinsky and
C.~F.~Steinwachs,
  arXiv:0910.1041 [hep-ph].


\bibitem{Okada:2009wz}
  N.~Okada, M.~U.~Rehman and Q.~Shafi,
  arXiv:0911.5073 [hep-ph].  


\bibitem{Bezrukov:2010jz}
  F.~Bezrukov, A.~Magnin, M.~Shaposhnikov and S.~Sibiryakov,
[arXiv:1008.5157 [hep-ph]].  

\bibitem{Lerner:2011it}
  R.~N.~Lerner and J.~McDonald,
arXiv:1112.0954 [hep-ph].  




\bibitem{Salopek:1988qh}
  D.~S.~Salopek, J.~R.~Bond and J.~M.~Bardeen,
Phys.\ Rev.\ D {\bf 40} (1989) 1753.  


\bibitem{Zee:1978wi}
  A.~Zee,
Phys.\ Rev.\ Lett.\  {\bf 42} (1979) 417.  


\bibitem{Adler:1982ri}
  S.~L.~Adler,
Rev.\ Mod.\ Phys.\  {\bf 54} (1982) 729   
[Erratum-ibid.\  {\bf 55} (1983) 837].  

\bibitem{Accetta:1985du}
  F.~S.~Accetta, D.~J.~Zoller and M.~S.~Turner,
Phys.\ Rev.\ D {\bf 31} (1985) 3046.  


\bibitem{CervantesCota:1994zf}
J.~L.~Cervantes-Cota and H.~Dehnen,
Phys.\ Rev.\ D {\bf 51} (1995) 395  [astro-ph/9412032].  


\bibitem{Kaloper:2008gs}
N.~Kaloper, L.~Sorbo and J.~'i.~Yokoyama,
Phys.\ Rev.\ D {\bf 78} (2008) 043527  [arXiv:0803.3809 [hep-ph]].  

\bibitem{Shaposhnikov:2008xb}
  M.~Shaposhnikov and D.~Zenhausern,
Phys.\ Lett.\ B {\bf 671} (2009) 187  [arXiv:0809.3395 [hep-th]].  

\bibitem{Shaposhnikov:2008xi}
  M.~Shaposhnikov and D.~Zenhausern,
Phys.\ Lett.\ B {\bf 671} (2009) 162  [arXiv:0809.3406 [hep-th]].  


\bibitem{Blas:2011ac}
  D.~Blas, M.~Shaposhnikov and D.~Zenhausern,
Phys.\ Rev.\ D {\bf 84} (2011) 044001  [arXiv:1104.1392 [hep-th]].  

\bibitem{GarciaBellido:2011de}
  J.~Garc\'{\i}a-Bellido, J.~Rubio, M.~Shaposhnikov and D.~Zenhausern,
Phys.\ Rev.\ D {\bf 84} (2011) 123504  [arXiv:1107.2163 [hep-ph]].  

\bibitem{grs}
  
J.~Garc\'{\i}a-Bellido, J.~Rubio and M.~Shaposhnikov,
arXiv:1209.2119 [hep-ph].  




\bibitem{Einhorn:2009bh}
  M.~B.~Einhorn and D.~R.~T.~Jones,
JHEP {\bf 1003} (2010) 026  [arXiv:0912.2718 [hep-ph]].  

\bibitem{Ferrara:2010yw}
  S.~Ferrara, R.~Kallosh, A.~Linde, A.~Marrani and A.~Van Proeyen,
Phys.\ Rev.\ D {\bf 82} (2010) 045003  
[arXiv:1004.0712 [hep-th]].  


\bibitem{Lee:2010hj}
  H.~M.~Lee,
JCAP {\bf 1008} (2010) 003  
[arXiv:1005.2735 [hep-ph]].  

\bibitem{Ferrara:2010in}
  S.~Ferrara, R.~Kallosh, A.~Linde, A.~Marrani and A.~Van Proeyen,
Phys.\ Rev.\ D {\bf 83} (2011) 025008  [arXiv:1008.2942 [hep-th]].  

\bibitem{Nakayama:2010sk}
  K.~Nakayama and F.~Takahashi,
JCAP {\bf 1102} (2011) 010  [arXiv:1008.4457 [hep-ph]].


\bibitem{BenDayan:2010yz}
  I.~Ben-Dayan and M.~B.~Einhorn,
JCAP {\bf 1012} (2010) 002  
[arXiv:1009.2276 [hep-ph]].  


\bibitem{Nakayama:2010ga}
  K.~Nakayama and F.~Takahashi,
JCAP {\bf 1011} (2010) 039  
[arXiv:1009.3399 [hep-ph]].  

\bibitem{Kallosh:2010xz}
  R.~Kallosh, A.~Linde and T.~Rube,
Phys.\ Rev.\ D {\bf 83} (2011) 043507  
[arXiv:1011.5945 [hep-th]].  





\bibitem{Pallis:2011ps}
C.~Pallis and N.~Toumbas,
JCAP {\bf 1102} (2011) 019  [arXiv:1101.0325 [hep-ph]].  


\bibitem{Arai:2011nq}
  M.~Arai, S.~Kawai and N.~Okada,
Phys.\ Rev.\ D {\bf 84} (2011) 123515  
[arXiv:1107.4767 [hep-ph]].  

\bibitem{Nakayama:2011ri}
  K.~Nakayama and F.~Takahashi,
[arXiv:1108.0070 [hep-ph]].  


\bibitem{Pallis:2011gr}
C.~Pallis and N.~Toumbas,
JCAP {\bf 1112} (2011) 002  [arXiv:1108.1771 [hep-ph]].  



\bibitem{Li:1973mq}
  L.~-F.~Li,
  Phys.\ Rev.\ D {\bf 9} (1974) 1723.



\bibitem{Georgi:1974sy}
  H.~Georgi and S.~L.~Glashow,
Phys.\ Rev.\ Lett.\  {\bf 32} (1974) 438.  

\bibitem{Feldmann:2010yp}
 T.~Feldmann,
JHEP {\bf 1104} (2011) 043  [arXiv:1010.2116 [hep-ph]].  

\bibitem{Kannike:2011fx}
K.~Kannike and D.~V.~Zhuridov,
JHEP {\bf 1107} (2011) 102  [arXiv:1105.4546 [hep-ph]].  


\bibitem{Marzocca:2011dh}
  D.~Marzocca, S.~T.~Petcov, A.~Romanino and M.~Spinrath,
JHEP {\bf1111} (2011) 009  [arXiv:1108.0614 [hep-ph]].  

\bibitem{Schnitter:2012bz}
  K.~Schnitter,
arXiv:1204.2111 [hep-ph].  


\bibitem{Wess:1992cp}
  J.~Wess and J.~Bagger, 
``Supersymmetry and supergravity,''  Princeton, USA: Univ. Pr. (1992) 259 p

\bibitem{Witten}
  E.~Witten,
  Phys.\ Lett.\  B {\bf 105} (1981) 267.
\bibitem{Einhorn:1982pp}
  M.B.~Einhorn and D.R.T.~Jones,
  Nucl.\  Phys.\ B  {\bf 211} (1983) 29.
\bibitem{Dimopoulos:1982gm}
  S.~Dimopoulos and S.~Raby,
Nucl.\ Phys.\ B {\bf 219} (1983) 479.  
\bibitem{Yamagishi:1982hy}
  H.~Yamagishi,
Nucl.\ Phys.\ B {\bf 216} (1983) 508.  


\end{thebibliography}
\end{document}